\documentclass{article}
\usepackage{spconf}

\usepackage{amsmath, amssymb, amsfonts}
\usepackage{algorithmic}
\usepackage{graphicx}
\usepackage{makecell}
\usepackage{hyperref}
\hypersetup{hidelinks}
\usepackage{textcomp}
\usepackage{xcolor}
\usepackage{booktabs}
\usepackage{multirow}
\usepackage{array}

\usepackage{adjustbox}
\usepackage{kotex}
\usepackage{subcaption}
\usepackage{caption}
\ninept

\begin{document}
\title{KOMOS: Automatic Mean Opinion Score Estimation for Korean Speech
}

\name{Sangchun Ha$^{1}$, 
Junwon Moon$^{2}$,
Kyuhong Shim$^{2}$}
\address{$^{1}$Independent Researcher \quad
    $^{2}$ Sungkyunkwan University, Republic of Korea \\
    \texttt{\small seomk9896@gmail.com, \{mppn98, khshim\}@skku.edu}}

\maketitle

\begin{abstract}
    As speech generation and real-time spoken dialogue systems advance, neural mean opinion score (MOS) predictors are increasingly used to evaluate speech quality.
However, many existing predictors are trained primarily on English read speech and generalize inconsistently across languages and speaking styles.
In particular, their reliability on Korean speech has not been adequately validated.
We propose KOMOS, a Korean MOS predictor for evaluating the overall quality of both read and spontaneous speech.
To train and evaluate KOMOS, we collect 5,000 ratings from 29 native Korean listeners for 1,000 utterances spanning real recordings, synthesized speech, and degraded speech.
KOMOS achieves an utterance-level Spearman rank correlation (SRCC) of 0.900, compared with 0.703 for UTMOS v1 and 0.767 for UTMOS v2.
The inference code is publicly available at \href{https://github.com/upskyy/komos}{\raisebox{-0.2\height}{\includegraphics[height=1.2em]{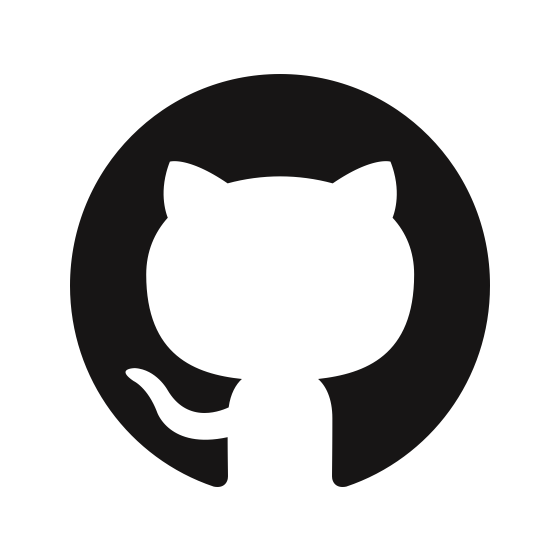}}}\,\url{https://github.com/upskyy/komos} and the trained model at \href{https://huggingface.co/upskyy/komos}{\raisebox{-0.2\height}{\includegraphics[height=1.0em]{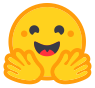}}}\,\url{https://huggingface.co/upskyy/komos}.

\end{abstract}


\section{Introduction}\label{sec:intro}

Recent advances in text-to-speech (TTS) and spoken dialogue systems have increased the need to evaluate perceived speech quality across both read and spontaneous speech~\cite{casanova24_interspeech, lee2025hierspeech++, lu2025slide}.
Subjective mean opinion score (MOS) tests remain the standard for speech quality assessment, but collecting human ratings is costly and time-consuming.
Neural MOS predictors therefore serve as practical tools for scalable evaluation. 
In particular, UTMOS v1~\cite{saeki22c_interspeech} is widely used as an automatic evaluation metric for zero-shot TTS systems.

\begin{figure}[t]
    \centering
    \includegraphics[width=1.0\linewidth]{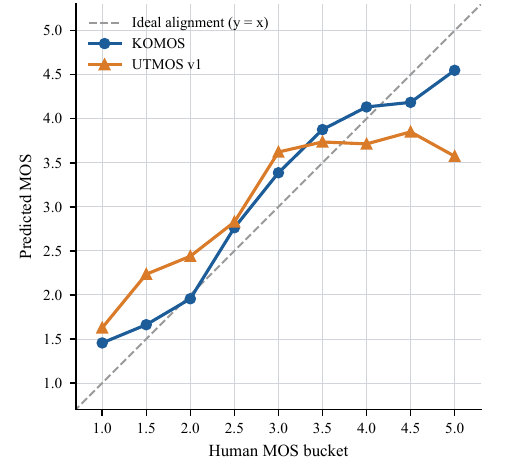}
    \caption{Predicted mean opinion score of KOMOS and UTMOS v1 within each human-MOS bucket. Predictions are affine-calibrated on the validation set; the dashed line indicates perfect alignment. KOMOS follows the human MOS scale more closely.}
    \label{fig:mos_alignment}
\end{figure}

However, many existing MOS predictors are trained primarily on English read speech from datasets such as BVCC~\cite{cooper21_ssw}, and prior studies report inconsistent generalization across languages and domains~\cite{cooper2022generalization, cooper2023voicemos, minixhofer2026ttsds2}.
UTMOS has also been used to evaluate Korean TTS systems~\cite{kwon25_interspeech,kwon2026coreaspeech}, although its reliability on Korean speech has not been adequately validated against native Korean listeners' judgments.
For example, our analysis shows that the calibrated mean predictions of UTMOS v1 overestimate human MOS in the 1.0--2.0 range and underestimate it in the 4.0--5.0 range (Figure~\ref{fig:mos_alignment}).
These discrepancies highlight the need to align MOS predictions with Korean listeners' judgments across speaking styles and quality levels.

To address this problem, we propose KOMOS, a Korean MOS predictor trained on human speech quality ratings covering both read and spontaneous speech. 
We fine-tune a pretrained speech encoder and train a listener-conditioned prediction head using human ratings of Korean speech.
To train and evaluate KOMOS, we construct a dataset of 1,000 utterances with 5,000 ratings from 29 native Korean listeners. 
The dataset includes real recordings, synthesized speech, and degraded speech to cover varied recording conditions and quality levels.
We compare multiple pretrained backbones and fine-tuning strategies under a common evaluation protocol to select the final KOMOS configuration.

KOMOS achieves an utterance-level test Spearman rank correlation coefficient (SRCC) of 0.900 and linear correlation coefficient (LCC) of 0.950, outperforming UTMOS v1 and v2 on both metrics. 
The SRCC improvements also extend to speaker-disjoint and text-disjoint subsets of the test set.
Beyond correlation, Figure~\ref{fig:mos_alignment} shows that KOMOS follows the human MOS scale more closely than UTMOS v1, including in the high-MOS range. 
In a separate diagnostic using speech from an unseen TTS system, KOMOS's calibrated mean predictions across all ten categories are consistent with the high-quality screening criterion (Sec.~\ref{ssec:exp_category}).
Together, these results support the value of Korean human ratings for adapting MOS predictors and demonstrate improved alignment with human judgments in both relative quality rankings and absolute MOS values.

Our contributions are as follows:
\begin{itemize}
    \item We propose KOMOS, a Korean MOS predictor for both read and spontaneous speech. 
    \item We construct a Korean MOS dataset comprising 1,000 utterances and 5,000 ratings from 29 native Korean listeners, covering diverse speech sources, styles, and quality levels.
    \item We analyze the limitations of existing MOS predictors on Korean speech and systematically compare pretrained backbones and fine-tuning strategies, demonstrating improved agreement between KOMOS predictions and human ratings.
  
\end{itemize}

\section{Related Work}\label{sec:related}

\subsection{Neural MOS Prediction and Generalization}\label{ssec:related_mos}

Non-intrusive MOS prediction has progressed from MOSNet~\cite{lo19_interspeech} to listener-aware models such as MBNet~\cite{leng2021mbnet} and LDNet~\cite{huang2022ldnet}, as well as regression using self-supervised speech features in SSL-MOS~\cite{cooper2022generalization}.
UTMOS v1~\cite{saeki22c_interspeech} combines listener embeddings with contrastive learning and achieved leading performance on multiple metrics in the VoiceMOS Challenge 2022.
UTMOS v2~\cite{baba2024t05} additionally adopts feature fusion with multi-stage training and attained top-tier performance in the 2024 VoiceMOS Challenge.
Training resources remain primarily English-centered, including BVCC~\cite{cooper21_ssw} and SOMOS~\cite{maniati22_interspeech}.
SQuId~\cite{sellam2023squid} demonstrates multilingual MOS prediction, using a large in-house human-rating corpus that is not publicly released.
IndicMOS~\cite{udupa24b_interspeech} also extends MOS prediction beyond English.
However, cross-lingual analyses of SSL-MOS and VoiceMOS Challenge tracks report inconsistent generalization~\cite{cooper2022generalization, cooper2023voicemos}, and MOS-Bench~\cite{huang2026mos} considers out-of-distribution generalization as a distinct challenge.
The reliability of existing predictors on Korean speech across speaking styles and quality levels remains insufficiently characterized.

\subsection{Korean Speech Quality Assessment}\label{ssec:related_korean}

Korean prosody is characterized by tonal patterns at the accentual phrase level~\cite{jun1998accentual}.
Korean speech also exhibits phonological processes such as liaison and tensification~\cite{kim1998unlimited, lee09c_interspeech}.
Recent Korean TTS studies employ UTMOS as an automatic evaluation metric; PEFT-TTS reports UTMOS scores for both single- and multi-speaker Korean speech synthesis~\cite{kwon25_interspeech}, and CoreaSpeech uses the same metric for both TTS evaluation and quality-based filtering of Korean speech data~\cite{kwon2026coreaspeech}. 
Emilia~\cite{he2024emilia}, a multilingual speech corpus that includes Korean, likewise uses an automated speech quality assessment tool to filter recordings collected from diverse online sources.
By learning from native Korean listeners’ overall quality ratings, KOMOS may also support quality-based selection of Korean speech data across speaking styles.

\section{KOMOS: A Korean MOS Predictor}\label{sec:method}

\subsection{Motivation}\label{ssec:method_motivation}

\begin{figure}[t]
    \centering
    \includegraphics[width=\columnwidth]
    {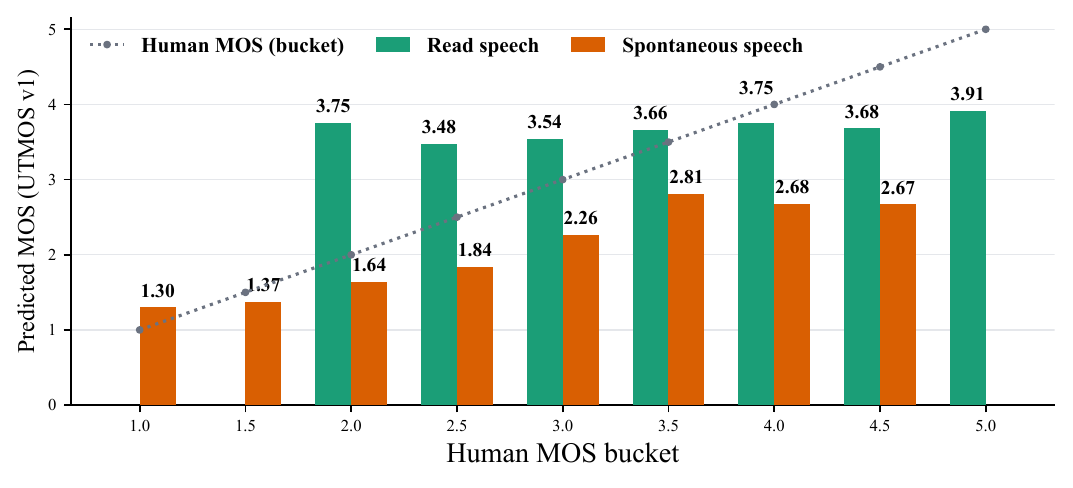}
    \caption{UTMOS v1 consistently underestimates spontaneous Korean speech relative to read speech within the same human-MOS buckets. Bars show the mean raw UTMOS v1 output within each human-MOS bucket, and the dotted line is the corresponding human rating.}
    \label{fig:speech_style_bias}
\end{figure}

To motivate the inclusion of spontaneous speech in Korean MOS prediction, we examine whether UTMOS v1 assigns comparable scores to read and spontaneous speech within similar human MOS ranges (Fig.~\ref{fig:speech_style_bias}).
We group utterances by rounding their human MOS to the nearest 0.5.
We compare the four buckets centered at 3.0, 3.5, 4.0, and 4.5, each containing at least five utterances of each speaking style.
The numbers of read/spontaneous utterances are 21/19, 49/10, 125/10, and 36/5, respectively.

Within these buckets, UTMOS v1 scores spontaneous speech 0.85--1.28 points below read speech, with an unweighted mean gap of 1.05 points across four buckets.
All four comparisons yield unadjusted $p<.001$ under two-sided Mann--Whitney U tests and remain significant after Bonferroni correction.
Because this analysis is not a paired experiment controlling for sentence and speaker, the observed differences cannot be attributed solely to speaking style.

\subsection{Dataset Construction}\label{ssec:method_dataset}
We construct a dataset of 1,000 Korean utterances comprising 420 real recordings (42\%), 310 TTS utterances (31\%), and 270 degraded utterances (27\%).
The audio includes studio read speech, spontaneous and noisy real recordings, diverse commercial and open-source TTS systems, and vocoder- or codec-degraded speech.
To cover various utterance types and Korean linguistic phenomena, we organize the dataset into six general categories---Command/\allowbreak Information (C), Free-talk (F), Question (Q), Emotion (E), Interjection (P), and Syntactic Complexity (S)---and four Korean-specific categories: tensification, liaison, Sino-Korean numeral readings, and honorifics.
The collection contains more than 200 human speakers and TTS voices.

In the dataset, the read/spontaneous distinction is based on linguistic style: read speech uses phrasing typical of written texts, whereas spontaneous speech uses conversational phrasing.
These categories describe the style of the utterance rather than whether it was produced without a script.

\subsection{Rating Protocol and Data Split}\label{ssec:method_rating}

Twenty-nine native Korean listeners rated overall speech quality from 1 to 5 through an in-house web interface.
Listeners were not shown transcripts and were instructed not to penalize speech merely for being formal or informal.
Five different listeners rated each utterance, yielding 5,000 ratings.
We define utterance-level MOS as the arithmetic mean of these five ratings.
Inter-rater reliability, measured by ordinal Krippendorff's $\alpha$, is $0.700$~\cite{krippendorff2019content}.

We stratify the dataset by utterance-level MOS and construct train, validation, and test splits of 810, 90, and 100 utterances, respectively.
The main split is a seen-system setting in which source domains and synthesis systems may occur across splits.
Within the test set, we additionally define a 34-utterance speaker-disjoint subset whose speakers do not occur in train or validation, and a 55-utterance text-disjoint subset whose sentences do not occur in either set.

\begin{table}[!t]
  \centering
  \caption{Utterance-level SRCC/LCC of zero-shot baselines and Korean-fine-tuned models on the validation and test sets. Bold indicates the best value in each metric column. ZS = zero-shot, FT = full fine-tuning, FE = frozen early.}
  \label{tab:backbone-method}
  \setlength{\tabcolsep}{4pt}
  \renewcommand{\arraystretch}{1.05}
  \resizebox{1.0\linewidth}{!}{
  \begin{tabular}{lrccccc}
    \toprule
    & & & \multicolumn{2}{c}{Validation} & \multicolumn{2}{c}{Test} \\
    \cmidrule(lr){4-5} \cmidrule(lr){6-7}
    Model & Params & Strategy & SRCC $\uparrow$ & LCC $\uparrow$ & SRCC $\uparrow$ & LCC $\uparrow$ \\
    \midrule
    \multicolumn{7}{l}{\emph{English-trained baselines (zero-shot)}} \\
    \addlinespace
    UTMOS v1                & 102.8M  & ZS & 0.734 & 0.807 & 0.703 & 0.798 \\
    UTMOS v2                & 203.7M  & ZS & 0.715 & 0.806 & 0.767 & 0.858 \\
    \midrule
    \multicolumn{7}{l}{\emph{KOMOS (Korean fine-tuned)}} \\
    \addlinespace
    UTMOS v1 init.          & 94.4M   & FT      & $0.920$ & $0.931$ & $0.865$ & $0.928$ \\
    \addlinespace
    wav2vec2-base           & 94.4M   & FE & 0.890 & 0.930 & 0.867 & 0.932 \\
    WavLM-base+             & 94.4M   & FT      & $0.899$ & $0.940$ & $0.885$ & $0.942$ \\
    XLS-R 300M              & 315.4M  & FE & $0.919$ & $0.943$ & $0.877$ & $0.941$ \\
    WavLM-large             & 315.5M  & FE & 0.922 & 0.945 & 0.895 & 0.944 \\
    \addlinespace
    Whisper-tiny            & 8.2M    & FT      & 0.901 & 0.917 & 0.859 & 0.925 \\
    Whisper-base            & 20.6M   & FT      & 0.926 & 0.940 & 0.874 & 0.933 \\
    Whisper-small           & 88.2M   & FT      & $0.922$ & $0.937$ & $0.895$ & $0.947$ \\
    Whisper-medium & 307.2M & FT & $\mathbf{0.932}$ & $\mathbf{0.949}$ & $0.900$ & $\mathbf{0.950}$ \\
    Whisper-large-v2        & 636.8M  & FT      & 0.926 & 0.941 & 0.909 & 0.948 \\
    Whisper-large-v3        & 637.0M  & FT      & 0.927 & 0.946 & \textbf{0.910} & 0.938 \\
    \bottomrule
  \end{tabular}}
\end{table}

\subsection{Model and Training}\label{ssec:method_model}

KOMOS follows the listener-conditioned model architecture of UTMOS~\cite{saeki22c_interspeech}: a pretrained speech encoder, a 128-dimensional listener embedding, and a regression head.
We compare multiple pretrained encoders and fine-tuning strategies and select the final configuration based on validation SRCC.
The selected model utilizes the Whisper-medium encoder~\cite{radford2023robust}, discarding its decoder and truncating its positional embedding table to 500 positions for 10-second inputs.
Audio is converted to 16-kHz mono.
We freeze the convolutional front end and fine-tune the remaining encoder parameters.

The listener embedding table contains 31 entries: 29 listeners and two aggregate entries, \textsc{mean} and \textsc{median}.
During training, each utterance is associated with $K=7$ targets: five individual quality ratings, their mean, and their median.
The encoder processes each utterance once, and its output is reused across targets.
For each target, the corresponding embedding is repeated over time and concatenated with each encoder frame.
The resulting sequence is passed through a single-layer bidirectional LSTM with a hidden size of 256, followed by a linear projection to a scalar score at each frame.
During inference, we use the \textsc{mean} embedding and average the scores over valid frames to obtain a single utterance-level MOS prediction.

We jointly optimize a frame-level regression loss for rating prediction and an utterance-level ranking loss for relative quality ordering.
Let $s_{ikf}$ be the head's score at valid frame $f$ for target $k$ of utterance $i$, $y_{ik}$ the corresponding target, and $F_i$ the number of valid frames after masking padding.
We use a frame-level clipped-MSE variant~\cite{saeki22c_interspeech,leng2021mbnet, huang2022ldnet},
\begin{align}
    \ell_{ik}
    &=\frac{1}{F_i}\sum_{f=1}^{F_i}
    \left[\max\!\left(\lvert s_{ikf}-y_{ik}\rvert-\tau,\,0\right)\right]^2,\\
    \mathcal{L}_{\mathrm{frame}}
    &=\frac{1}{BK}\sum_{i=1}^{B}\sum_{k=1}^{K}\ell_{ik},
    \label{eq:frame_loss}
\end{align}
where $B$ is the batch size and $\tau=0.25$ defines an error tolerance.
Errors within this threshold incur no penalty, while larger errors are penalized by their squared excess over the threshold.
We average this loss over valid frames, $K$ targets, and $B$ utterances in the batch. 

For the ranking loss, we average the frame scores conditioned on the \textsc{mean} embedding over valid frames.
We denote these utterance-level predictions by $\hat{y}_i$ and the corresponding mean human ratings by $y_i$.
Within each batch, we select utterance pairs whose mean human ratings differ by at least $\delta=0.5$, forming a set of pairs $\mathcal{P}=\{(i,j):i<j,\ \lvert y_j-y_i\rvert\geq\delta\}$. 
The ranking loss is
\begin{equation}
    \mathcal{L}_{\mathrm{rank}}
    =\frac{1}{\lvert\mathcal{P}\rvert}
    \sum_{(i,j)\in\mathcal{P}}
    \max\!\left(0,\,m-\sigma_{ij}\!(\hat{y}_j-\hat{y}_i)\right),
    \label{eq:rank_loss}
\end{equation}
where $\sigma_{ij}=\operatorname{sgn}(y_j-y_i)$ and $m=0.1$.
A pair incurs no loss once its predictions have the correct order with a margin of at least $m$.
The final objective is $\mathcal{L}=\mathcal{L}_{\mathrm{frame}}+\lambda\mathcal{L}_{\mathrm{rank}}$, with $\lambda=0.5$.

We train KOMOS using AdamW with learning rates of $1\times10^{-5}$ for the encoder and $3\times10^{-4}$ for the head and embeddings.
We use weight decay of 0.01, batch size of 32, and bf16 precision.
Training runs for up to 20 epochs, with early stopping after five consecutive epochs without improvement in validation SRCC.
For absolute-score analyses, we fit an affine calibration $\hat{y}_{\text{cal}}=a\hat{y}+b$, from model predictions to human MOS on the validation set, and clip the calibrated predictions to the range $[1,5]$.
The calibration procedure is applied separately to each predictor included in these analyses.

\begin{table}[!t]
  \centering
  \caption{Utterance-level SRCC and LCC for four training strategies using the Whisper-medium encoder.}
  \label{tab:whisper-medium-strategy}
  \setlength{\tabcolsep}{13pt}
  \renewcommand{\arraystretch}{1.05}
  \resizebox{1.0\linewidth}{!}{
  \begin{tabular}{lcccc}
    \toprule
    & \multicolumn{2}{c}{Validation} & \multicolumn{2}{c}{Test} \\
    \cmidrule(lr){2-3} \cmidrule(lr){4-5}
    Strategy & SRCC $\uparrow$ & LCC $\uparrow$ & SRCC $\uparrow$ & LCC $\uparrow$ \\
    \midrule
    \textbf{full-FT} & $\mathbf{0.932}$ & $\mathbf{0.949}$ & $\mathbf{0.900}$ & $\mathbf{0.950}$ \\
    frozen-ws        & 0.907 & 0.925 & 0.871 & 0.936 \\
    frozen-last      & 0.911 & 0.927 & 0.860 & 0.924 \\
    frozen-early     & 0.888 & 0.909 & 0.840 & 0.912 \\
    \bottomrule
  \end{tabular}}
\end{table}

\begin{table}[!t]
  \centering
  \caption{Utterance-level SRCC on the full test set and its speaker-disjoint ($n=34$) and text-disjoint ($n=55$) subsets. Subset speakers or texts are absent from both training and validation.}
  \label{tab:disjoint}
  \footnotesize
  \setlength{\tabcolsep}{4pt}
  \renewcommand{\arraystretch}{1.05}
  \begin{tabular}{lccc}
    \toprule
    Model & \shortstack{Overall\\($n=100$)} & \shortstack{Speaker-disjoint\\($n=34$)} & \shortstack{Text-disjoint\\($n=55$)} \\
    \midrule
    UTMOS v1 (zero-shot) & 0.703 & 0.818 & 0.719 \\
    UTMOS v2 (zero-shot) & 0.767 & 0.823 & 0.778 \\
    \textbf{KOMOS} & \textbf{0.900} & \textbf{0.951} & \textbf{0.878} \\
    \bottomrule
  \end{tabular}
\end{table}

\begin{table*}[!t]
  \centering
  \caption{Mean calibrated predictions for 100 utterances synthesized from an unseen TTS system. Each utterance was selected by human-MOS of at least 4.5. The set covers ten categories ($n = 10$ each), with one representative utterance shown per category.}
  \label{tab:category-bonus}
  \footnotesize
  \setlength{\tabcolsep}{4pt}
  \renewcommand{\arraystretch}{1.1}
  \begin{tabular}{@{}l p{3.6cm} p{6.0cm} ccc@{}}
    \toprule
    Category & Description & Example utterance & UTMOS v1 & UTMOS v2 & KOMOS \\
    \midrule
    C (Command/Info) & Commands and factual information delivery
      & 오늘 회의는 오후 3시에 시작됩니다. & 3.075 & 3.102 & 4.743 \\
    F (Free-talk) & Casual everyday conversation
      & 발코니에 캠핑 의자를 놓았어요. & 3.055 & 3.293 & 4.616 \\
    Q (Question) & Interrogatives carrying question intonation
      & 이 근처에 편의점 있나요? & 3.226 & 3.300 & 4.675 \\
    E (Emotion) & Emotionally colored speech
      & 드디어 시험이 끝나서 너무 홀가분해요. & 2.924 & 3.426 & 4.750 \\
    P (Interjection) & Utterance-initial interjections and discourse markers
      & 아이고, 또 늦잠을 자 버렸어요. & 3.248 & 3.520 & 4.665 \\
    S (Syntactic complexity) & Long, syntactically complex sentences (avg. 12.8 words)
      & 아침부터 내리던 비가 오후가 되어서야 그치고 하늘이 서서히 맑아지기 시작했다. & 2.864 & 3.343 & 4.742 \\
    K-Tensification & Post-obstruent tensification
      & 국밥[국빱] 대신 국수[국쑤]를 주문했어요. & 3.116 & 3.377 & 4.674 \\
    K-Liaison & Resyllabification across a syllable boundary
      & 꽃이[꼬치] 활짝 피었어요. & 3.429 & 3.604 & 4.673 \\
    K-Numerals & Sino-Korean numeral readings from Arabic-digit expressions
      & 회의실은 3층[삼 층]에 있습니다. & 3.232 & 2.954 & 4.763 \\
    K-Honorifics & Subject-honorific and humble verb forms
      & 잠시만 기다려 주시면 안내해 드릴게요. & 3.120 & 3.496 & 4.714 \\
    \midrule
    \textbf{Overall mean} & & & 3.129 & 3.342 & \textbf{4.702} \\
    \bottomrule
  \end{tabular}
\end{table*}

\section{Experiments}\label{sec:experiment}

\subsection{Setup}\label{ssec:exp_setup}
We evaluate utterance-level predictions using Spearman’s rank correlation coefficient (SRCC) and Pearson’s linear correlation coefficient (LCC), with SRCC as the primary metric.
All training experiments use three random seeds, and we report the mean of the metrics computed separately for each seed.
Checkpoints and the final configuration are selected by validation SRCC.
We use public UTMOS v1 and UTMOS v2 checkpoints as zero-shot baselines without parameter updates on Korean data.

For Korean adaptation, we compare wav2vec2-base, WavLM-base+, WavLM-large, XLS-R-300M, and Whisper encoders from tiny to large-v3.
``UTMOS v1 init.'' denotes initialization with encoder weights from the released UTMOS v1 checkpoint.
All adapted models share the same prediction head architecture and training objectives.
Parameter counts refer to the encoder for adapted models and the complete predictor for the baselines.

We compare four training strategies. \emph{Full-FT} (FT in Table~\ref{tab:backbone-method}) updates the encoder, prediction head, and listener embeddings, except for Whisper's frozen convolutional front end.
The other strategies keep the encoder fixed while training the head and listener embeddings.
\emph{Frozen-ws} combines the outputs of all encoder layers by linear combination using learnable weights, whereas \emph{frozen-last} uses only the final-layer output.
\emph{Frozen-early} (FE) uses the output of the single layer with the largest learned weight in the corresponding \emph{frozen-ws} model.

\subsection{Korean MOS Prediction and Model Selection}\label{ssec:exp_main}
Table~\ref{tab:backbone-method} reports the strategy with the highest validation SRCC for each KOMOS backbone. 
First, all backbones adapted using Korean ratings outperform both ZS baselines, including Whisper-tiny, whose encoder is an order of magnitude smaller than the baselines.
This indicates that model capacity alone does not explain the gap and highlights the value of in-language supervision.
Second, Whisper-medium achieves the highest validation SRCC among all evaluated backbones and is selected as the final backbone.
Although larger Whisper variants obtain slightly higher test SRCCs, they have lower validation SRCCs and require more parameters.
For Whisper-medium, full fine-tuning outperforms all three frozen-encoder strategies on both metrics (Table~\ref{tab:whisper-medium-strategy}).

The selected KOMOS configuration achieves a test SRCC of 0.900 and LCC of 0.950. 
A paired bootstrap over the 100 test utterances with 10,000 resamples, averaging SRCC across three seeds within each resample, yields 95\% confidence intervals for the SRCC improvement of 0.100--0.320 over UTMOS v1 and 0.056--0.231 over UTMOS v2 (both p $<$ 0.001).

\subsection{Speaker- and Text-Disjoint Evaluation}\label{ssec:exp_disjoint}

KOMOS achieves higher SRCC than both baselines on the speaker- and text-disjoint subsets, achieving 0.951 and 0.878, respectively (Table~\ref{tab:disjoint}).
All three predictors show an increase in SRCC on the speaker-disjoint subset relative to the full test set, suggesting that subset composition may contribute to this increase.
These results support generalization to held-out speakers and texts within the main seen-system setting.
Note that overall SRCC is not a weighted average of subset SRCCs and can therefore be lower than both.

\subsection{Score Alignment on an Unseen TTS System}\label{ssec:exp_category}

Correlation alone does not establish alignment with the absolute human MOS scale.
We therefore conduct a diagnostic evaluation on 100 utterances synthesized with a single fixed voice from an unseen Korean TTS system that does not appear during training.
The set contains ten utterances per category (Sec.~\ref{ssec:method_dataset}), each filtered to meet a score threshold of 4.5. 
This experiment shows whether the score-placement pattern is confined to a particular category. 
Bracketed forms in Table~\ref{tab:category-bonus} indicate surface pronunciations.

After validation-based affine calibration, KOMOS assigns a mean score of at least 4.5 to all ten categories, whereas neither baseline reaches this threshold in any category. 
Its overall mean prediction is 4.702, compared with 3.129 for UTMOS v1 and 3.342 for UTMOS v2.

\section{Conclusion}\label{sec:conclusion}

We proposed KOMOS, a Korean MOS predictor trained on human quality ratings covering both read and spontaneous speech. 
To train KOMOS, we compared multiple pretrained speech encoders and fine-tuning strategies.
Using 5,000 ratings from 29 native Korean listeners for 1,000 utterances, KOMOS achieved a test SRCC of 0.900 and LCC of 0.950, outperforming UTMOS v1 and v2. 
KOMOS also showed closer alignment with the human MOS scale than UTMOS v1. 
In the unseen-system diagnostic, the model's mean predictions across all ten categories were consistent with the human annotators' high-quality screening criterion. 
These results highlight the value of in-language supervision for Korean MOS prediction. 
We publicly released the inference code and trained model to support Korean speech quality assessment.

\section{Compliance with Ethical Standards}

The listening study was conducted with adult participants who voluntarily provided quality ratings of speech samples on a 1--5 scale. No sensitive personal information was collected, and no personally identifiable participant information was reported or released.

\newpage
\bibliographystyle{IEEEbib}
\bibliography{reference}
\end{document}